# Mean shape of the human limbus

Alejandra Consejo[1], Clara Llorens-Quintana[1], Hema Radhakrishnan[2] and D. Robert Iskander[1]

[1]Department of Biomedical Engineering, Wroclaw University of Science and Technology

[2] Division of Pharmacy and Optometry, Faculty of Biology, Medicine and Health, The University of Manchester

Corresponding author: Alejandra Consejo, MSc, Department of Biomedical Engineering, Faculty of Fundamental Problems of Technology, Wroclaw University of Science and Technology, Wybrzeze Wyspianskiego 27, 50-370 Wroclaw, Poland. E-mail: alejandra.consejo@pwr.edu.pl.

This project has received funding from by the European Marie Curie ITN grant, AGEYE, 608049, and the Horizon 2020 research and innovation program under the Marie Sk1odowska-Curie grant, EDEN, 64276 (both public funding).

**ABSTRACT**

Purpose: To characterize the mean topographical shape of the human limbus of a normal eye and ascertain whether it depends on age and refractive power.

Setting: Academic institution.

Design: Prospective case series.

Methods: Seventy-four subjects aged from 20 to 84 years and with no previous ocular surgeries were included in this study. The left eye was measured four times with a corneo-scleral topographer (Eye Surface Profiler). From the raw anterior eye height data of each measurement, topographical limbus was demarcated and fitted in three dimensions to a circle, an ellipse and a Fourier series. Root mean square error (RMSE) was calculated to evaluate the goodness of fit. In addition, white-to-white (WTW) corneal diameter was taken from the readings of the measuring device and compared with the topographical limbus. For statistical analysis, subjects were grouped as young (< 35 y.o.) and older (> 50 y.o.), and also according their equivalent sphere correction.

Results: From the considered models, second-order Fourier series resulted in the most accurate model to describe the shape of the human limbus. The difference between the topographical limbus diameter and the WTW corneal diameter amounted on average to 0.33 ± 0.24 mm. Statistically significant differences among eye quadrants (P<0.001) were found. No statistically significant difference in horizontal and vertical meridian between age groups (P=0.71 and P=0.082, respectively) or between myopes and emmetropes (P=0.78 and P=0.68, respectively) were found.

Conclusions: Human limbus is not symmetrical and its shape is subject dependent but not related to age and eye's refractive power.

**SYNOPSIS**

The mean shape of the topographical normal limbus is presented. Normal limbus is not symmetrical, and is not related to age and the mean spherical equivalent refractive power of the eye.

**INTRODUCTION**

The cornea and the sclera are the main structures that constitute the anterior eye surface. At the junction of these two surfaces there is a change in the curvature which can be identified as the limbus. The anterior limbus is formed by the junction of the corneal and conjunctival epithelia where the epithelium gradually becomes thicker toward the sclera. Anatomically, limbus is of interest because it is a site of many different but important features related to eye functions and contains corneal stem cell for nourishing the cornea. It also contains the principal conventional pathways of aqueous humour outflow, needed, among other functions, for the maintenance of the intraocular pressure.[1]

The corneal diameter is defined as the limbus-to-limbus distance and clinically both the horizontal and vertical dimensions are regarded important. In ophthalmology, they are needed in the diagnosis of diseases such as macrocornea, microcornea, macrophtalmia, microphtalmia or relative anterior microphtalmia.[2] Delimiting corneal borders is also essential for refractive cataract surgeons as a useful approximation for sizing some types of anterior chamber intraocular lenses.[3]

White-to-white (WTW) corneal diameter, horizontal visible iris diameter (HVID) and vertical visible iris diameter (VVID) are the parameters usually used to estimate the corneo-scleral transition. Manual and automatized techniques coexist in the clinic to make these estimates. Table 1 summarizes, in a chronological order, the results of WTW corneal diameter estimates evaluated in previous studies.

Most of the estimates of the WTW corneal diameter presented in Table 1 are based on imaging the colored transition from the iris to the sclera. The estimates which considered topographical transition from cornea to sclera used manual calipers. However, it has been established that the corneo-scleral topographical transition does not necessary correspond to the color transition observable with en-face imaging.[23,24] In the past, the limitations of technology restricted the study of the anterior eye topography to the corneal region. Keratometers, keratoscopes, Scheimpflug-based topographers and optical coherence tomography (OCT) have been used to assess the shape of the human cornea.[25,26] On the other hand, the shape of the limbus, understood as the transition between cornea and sclera, has been conventionally considered circular. Although the importance of the topography of corneo-scleral transition has been recognized,[27] no accurate description of the human limbal shape has been reported. Nowadays it is possible to overcome the technical limitations that constrained the topographical studies to the corneal region, with non-contact commercially available instruments, such as corneo-scleral profilometers, that cover the corneo-scleral area far beyond the limbus.[28]

Table 1. Mean horizontal corneal diameter or WTW in different studies.

| Authors | Year | Number of eyes | Measuring device | WTW (mm) |
|---|---|---|---|---|
| **Martin & Holden** [4] | 1982 | 50 | Videotapes and photographs | 11.64 ± 0.49 |
| **Edmund** [5] | 1988 | 56 | Slit-lamp | 11.86 ± 0.55 |
| **Pop et al.** [6] | 2001 | 43 | Caliper | 11.87 ± 0.49 |
| **Baumesiter et al.** [7] | 2004 | 100 | Caliper<br>Holladay-Godwin gauge<br>Orbscan II<br>IOLMaster | 11.91 ± 0.71<br>11.80 ± 0.60<br>11.78 ± 0.43<br>12.02 ± 0.38 |
| **Werner et al.** [8] | 2004 | 12 (cadaver eyes) | Surgical caliper<br>Digital caliper | 11.77 ± 0.40<br>11.75 ± 0.43 |
| **Goldsmith et al.** [9] | 2005 | 40 | Holladay-Godwin gauge | 11.78 ± 0.57 |
| **Rüfer el at.** [10] | 2005 | 743 | Orbscan II | 11.71 ± 0.42 |
| **Kohnen et al.** [11] | 2006 | 52 | IOLMaster<br>Orbscan II | 12.17 ± 0.45<br>11.84 ± 0.41 |
| **Lim** [12] | 2006 | 724 | Orbscan II | 11.56 ± 0.36 |
| **Ronneburger et al.** [13] | 2006 | 65 (children) | Calipers together with microscope | 10.75 ± 0.81<br>10.65 ± 0.65 |
| **Piñero et al.** [14] | 2008 | 36 | Digital calliper (by CSO topographer) | 12.25 ± 0.49 |
| **Salouti et al.** [15] | 2009 | 74 | Galilei 4.01<br>Orbascan II | 12.01 ± 0.61<br>11.67 ± 0.29 |
| **Sanchis-Gimeno et al.** [16] | 2011 | 379 | Orbscan II | 11.9 ± 0.2 |
| **Nemeth et al.** [17] | 2010 | 91 | IOLMaster | 11.99 ± 0.47 |
| **Zha** [18] | 2012 | 231 (myopes) | Orbscan II | 11.49 ± 0.36 |
| **Martin et al.** [19] | 2013 | 328 | Orbscan II<br>IOLMaster | 11.69 ± 0.37<br>12.19 ± 0.40 |
| **Hall et al.** [20] | 2013 | 199 | Slit lamp | 11.7 ± 0.5 |
| **Domínguez-Vicent et al.** [21] | 2014 | 80 | Galilei G4<br>Pentacam HR | 11.84 ± 0.31<br>11.90 ± 0.43 |
| **Hickson-Curran et al.** [22] | 2014 | 255 | Digital camera | 11.75 ± 0.50 |
| **Chen et al.** [2] | 2015 | 729<br>977 | Calipers<br>IOLMaster | 12.22 ± 0.52<br>12.12 ± 0.42 |

The aim of this work was to characterize the mean topographical shape of the human limbus of a normal eye and ascertain whether it depends on age and refractive power.

**SUBJECTS AND METHODS**

Seventy-four participants were included in this study. They were adult subjects (47 females, 27 males) aged between 20 and 84 years (mean ± SD age: 33.1 ± 16.8). The study was approved by The University of Manchester Human Research Ethics Committee and adhered to the tenets of the Declaration of Helsinki. All subjects gave written informed consent to participate after the nature and possible consequences of the study were explained. All participants were free of ocular disease and current use of topical ocular medications was specified by the subjects as part of a background questionnaire. Exclusion criteria also included the presence of any corneal, conjunctiva or scleral pathology, any history or ocular surgery, as well as contact lens wear.

The study was performed in a single visit for each of the subjects. Firstly, the refractive power was measured in monocular conditions (the non-tested eye was occluded with an eye patch) using an open-view autorefractometer (Shin Nippon SRW-500, Ajinomoto Trading Inc., Japan). Subjects were asked to focus on a fixed 6-meter distant target (Maltese cross) positioned on a flat wall and the autorefractometer was aligned with the center of the pupil. Five measurements were acquired from the left eye of each subject. The autorefractometer provided the average value of those five measurements that we considered as the valid refractive sphero-cylindrical power for the eye under analysis.

Further, topographical data was obtained using a non-contact corneo-scleral topographer (Eye Surface Profiler (ESP), Eaglet Eye BV, Netherlands), a height profilometer with the potential to measure the corneo-scleral topography far beyond the limbus. To determine surface heights, algorithms used in ESP achieve similar levels of accuracy to those reached in keratoscopy based instruments such as Placido disk videokeratoscopes.[28] Fluorescein was instilled into the subject's eye. Dyed eye is needed for scleral data acquisition. Accurate measurements of anterior eye surface using ESP require instillation of fluorescein with a more viscous solution than saline.[28] The BioGlo (HUB Pharmaceuticals) ophthalmic strips were used to gently touch the upper temporal ocular surface. They were impregnated with 1 mg of fluorescein sodium ophthalmic moistened with one drop of an eye lubricant (HYLO-Parin, 1mg/ml of sodium hyaluronate). Four measurements were collected from the left eye of each subject. Subjects were instructed to open their eyes wide (not forcefully) prior the measurements with ESP to insure full coverage of the corneo-scleral area. Measurements in which corneo-scleral area was affected by eyelids were excluded.

Following data acquisition, the raw anterior eye height data (X, Y, and Z coordinates) was exported from ESP for further analysis. Limbal transition was calculated in 360 semi-meridians, using a custom written algorithm, as the point corresponding to a certain amount of change in the curvature between cornea and sclera.[23] Further, three different parametric models including a best-fit-circle (a three-parameter model), best-fit-ellipse (a five-parameter model) and a best-fit-Fourier-series were applied to limbal transition points demarcated earlier. The Fourier series is a sum of sine and cosine functions that describes a periodic signal (1),

$$y = a_0 + \sum_{i=1}^{n} a_i \cos(iwx) + b_i \sin(iwx) \quad (1)$$

where $a_0$ models a constant term in the data, $a_i$ and $b_i$ denote the Fourier coefficients, $w$ is the fundamental angular frequency of the signal and n is the number of terms in the series. We used the second order (n=2) Fourier series, because preliminary analysis showed that higher order series would not result in a model that has a statistically significant higher correlation coefficient (Fisher test, P>0.05). The second order Fourier series limits the model to two pairs of sin and cos functions and a constant, resulting in a six-parameter model ($a_0$, $a_1$, $a_2$, $b_1$, $b_2$, $w$).

Figure 1 shows an example of the limbal demarcation for a randomly chosen subject. Given the demarcated limbal points (indicated by small yellow overlapping circles), calculated using an appropriate algorithm,[23] for each acquired measurement, the three types of fit (circular, elliptical and Fourier series) were performed over each of the four measurements acquired per subject. The resulting fitting parameters were averaged per subject prior to statistical analysis. To compare the goodness of fit of the three models, the root mean square error (RMSE) was calculated, separately for each of the considered models. The RMSE of an estimator measures the average of the squares of the residuals between the estimates and raw data. In addition, WTW corneal diameter was taken from the readings of the measuring device (ESP) and compared with our results obtained from the raw *XYZ* anterior eye height data.

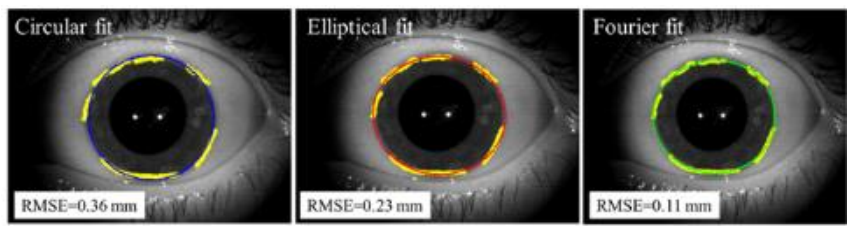

**Figure 1.** An example of the estimated limbus position indicated by small yellow overlapping circles (calculated using an automatic method described earlier[23]) and the fitting to the located points by a circle (blue solid line), an ellipse (red solid line), and Fourier series (green solid line), for a randomly selected subject (male, 21 years old). The goodness of the fit in each case is determined by the RMSE.

For statistical analysis, subjects were grouped according to their age. Forty-four subjects were grouped as young (under 35 years old) and 20 subjects were grouped as old (over 50 years old). In addition, subjects with less than 0.75 D of astigmatism were also grouped in two different categories according to their refractive power. 18 subjects were categorized as emmetropes (equivalent sphere correction between -0.50 D and +0.75 D) and 17 subjects were categorized as myopes (equivalent sphere correction between -1.50 D and -6.00 D).

The statistical analysis was performed using SPSS software for Windows version 23.0 (SPSS Inc., Chicago, Illinois, United States). The Shapiro-Wilk test was used to test the distribution type (Gaussian or non-Gaussian) of all continuous variables. The non-parametric Mann-Whitney test was used to ascertain whether there are limbal radial differences among quadrants, between young and older subjects and between myopes and emmetropes. Correlation of limbal radial distance and age and correlation of limbal radial distance and refractive power was explored using the Spearman's correlation coefficient (ρ). A probability of less than 5 % ($P < 0.05$) was considered statistically significant.

**RESULTS**

We found that the second-order Fourier series fit is the most appropriate (mean RMSE = 0.041 mm) to describe the shape of the human limbus, followed by elliptical fitting (mean RMSE = 0.15 mm) and circular fitting (mean RMSE = 0.56 mm). The mean human limbus is then described by a 2nd-order Fourier series (1) which parameters amounted on average to (95% confident bounds in brackets): $a_0$=6.059 (6.050, 6.069); $a_1$= -0.0710 (-0.101, -0.0412); $b_1$=-0.429 (-0.445, -0.412); $a_2$=0.132 (0.110, 0.154); $b_2$ =0.0289 (0.0115, 0.0462); w=0.0180 (0.0173, 0.0188). For a more simplistic approach we can consider the human limbus as an oblate ellipse, with an average major axis of 12.73 ± 0.41 mm, minor axis of 11.73 ± 0.49 mm and slightly tilted with respect to the horizontal meridian (1.03 ± 12.60°). Similarly, an even simpler approach is considering the limbus as circle. The limbal diameter from circular fitting amounted on average to 12.16 ± 0.35 mm, which coincides with the WTW readings of the ESP topographer (12.19 ± 0.26 mm). However, individual differences in absolute value between the circular fitting and the WTW reading amounted on average to 0.33 ± 0.24 mm. This difference was not statistically significant (Wilcoxon test, P = 0.60).

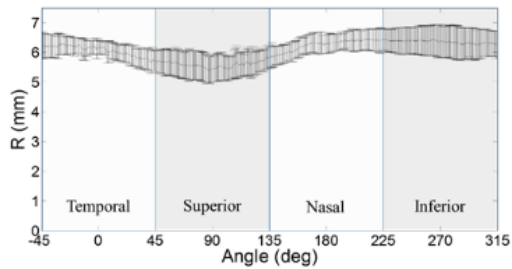

**Figure 2.** Mean radial distance in each sector of the eye, for 74 subjects. Error bars indicate +/- one standard deviation.

Limbus is not symmetrical. As Figure 2 shows the distance from the centre of the eye to the upper limbus is shorter than for the remaining sectors. Table 2 quantifies the averaged limbal radial distance per quadrant. The mean shape of the human limbus is shown in Figure 3. We found that the range in horizontal meridian amounts to [10.7, 14.2] mm and in vertical meridian to [10.5, 13.1] mm. More detailed information on the distribution of the horizontal and vertical radial distances is given in histograms in Figure 4.

**Table 2.** Mean radial distance and standard deviation in each quadrant, calculated from 74 eyes.

| Quadrant | Mean radial distance (mm) | Statistical significance (Mann-Whitney test) |
|---|---|---|
| Temporal | 6.07 ± 0.32 | P <<0.001 |
| Nasal | 6.25 ± 0.33 | |
| Superior | 5.60 ± 0.44 | P <<0.001 |
| Inferior | 6.36 ± 0.49 | |

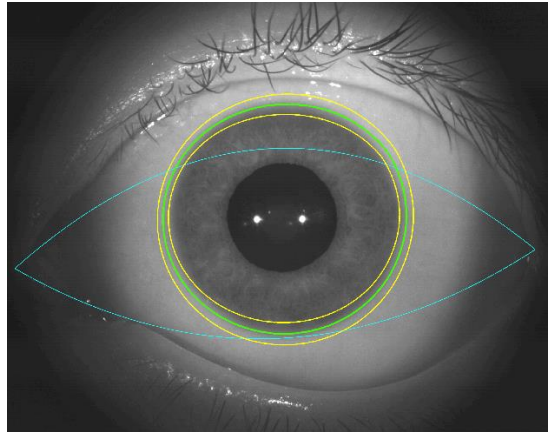

**Figure 3.** Model of the mean shape of the human topographical limbus (green line) and its range of variability calculated as one standard deviation from the mean (yellow lines), overlapping an image in which the eye is open sufficiently wide for the eyelids not to interfere with the estimated values. Cyan lines represent the average eyelids position in natural gaze position.29 Note that the superior part of the corneo-scleral junction is situated deeper under the upper eyelid than that located inferiorly.

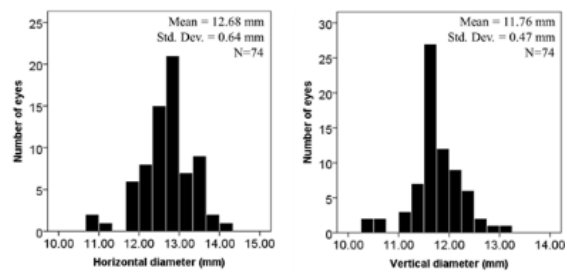

**Figure 4.** Population distribution of horizontal limbal diameter (left) and vertical limbal diameter (right) for 74 eyes.

Further, we investigated whether age-related differences in limbal shape exist, using Fourier series fitting, since it is the fitting with smallest error from the proposed ones. We analyzed the horizontal and vertical meridian of each subject and related it with their age. No correlation was found between horizontal meridian and age ($\rho=0.008$, P=0.95), vertical meridian and age ($\rho=-0.070$, P=0.55) or the ratio of horizontal and vertical meridian and age ($\rho=0.098$, P=0.41). Similarly, no statistical significant differences were found between young (under 35 years old) and old subjects (over 50 years old) in horizontal meridian (Mann-Whitney test, P=0.71), vertical meridian (Mann-Whitney test, P=0.082) or the ratio of horizontal and vertical meridian (Mann-Whitney test, P=0.18).

We also assessed whether there are differences in limbal shape depending on the refractive power. The spherical equivalent covered a range of [-6.0 D, +2.5 D]. No correlation was found between the horizontal meridian and refractive power ($\rho=0.12$, P =0.44), vertical meridian and refractive power ($\rho=0.075$, P=0.64) or the ratio of the horizontal and vertical meridians and refractive power (ρ=0.05, P=0.74). Likewise, no statistical significant differences were found between myopes and emmetropes in horizontal meridian (Mann-Whitney test, P=0.78), vertical meridian (Mann-Whitney test, P=0.68) or the ratio of the horizontal and vertical meridians (Mann-Whitney test, P=0.76).

Similarly, we did not find any statistically significant correlation between the cylinder power and the limbal radial distance in the meridian corresponding to the axis of the cylinder for the flattest meridian (ρ=-0.14, P=0.23) nor the steepest meridian (ρ=-0.11, P=0.34). No correlation was found either considering only those subjects with cylinder equal or greater than 0.75 D for the flattest meridian (ρ=0.080, P=0.69) nor the steepest meridian (ρ=0.22, P=0.26).

Statistical power post-hoc estimation was made. The analysis was conducted for 90% power at the 5% alpha level. For a sample size of 74 subjects, differences in limbal radius of about 140 micrometers could be differentiated.

**DISCUSSION**

From topographical corneo-scleral maps of seventy-four eyes we described, for the first time, the mean shape of the human limbus and we found that normal topographical limbus is not circular. In accordance with recent works[23,24] we found that the topographical limbus does not necessarily correspond to the WTW corneal diameter observable with enface imaging.

Three different types of fitting (circular, elliptical and Fourier series fit) were implemented to the demarcated points corresponding to the topographical transition between cornea and sclera calculated in 360 semi-meridians using a method described previously.[23] The complexity of the model and the accuracy of it are strongly linked. The Fourier series fit resulted in the most accurate (smallest RMSE) but the most complex (six parameters) model. Contrarily, the circular fitting was the simplest approach (three parameters) but it had the largest error. Additionally, it is worth pointing out that in some cases a substantial difference was found between the circular fitting estimated from topographical data and the WTW corneal diameter readings given by the measuring device (up to 1.2 mm). Hence, it is important to be aware of the limitations and differences among different methods of estimating the corneo-scleral transition.

We also found differences in radial distance among quadrants (see Figure 2 and Table 2). The superior semi-meridian was the shortest (5.60 ± 0.44) mm. This might be justified by the effect of the eyelid pressure on this area (see Figure 3). It is interesting to notice that the difference in radial distance in vertical meridian (both inferior and superior part) is more variable among subjects, standard deviation of about 0.5 mm, than the horizontal (both nasal and temporal part) meridian, standard deviation of about 0.3 mm. Indicating a higher subjective dependence on limbal shape in the vertical meridian than in the horizontal one.

Rüfer et al.[10] claimed that WTW corneal diameter decreases with age. However, we did not find statistically significant age-related differences in topographical limbal shape. Nevertheless, we found a correlation close to the significance (Mann-Whitney test, P=0.082) in the length of the vertical meridian between age groups. Older subjects had a slightly shorter vertical meridian (11.61 ± 0.42 mm), than the average value in the younger group (11.80 ± 0.46 mm). This slight difference might be justified by the effect of the eyelid pressure over time. Similarly, Martin el al.[19] claimed the myopic eye to have a smaller WTW corneal diameter than the emmetropic eye. However, we did not find any substantial differences in topographical limbal shape between myopes an emmetropes. These differences might be justified, once again, because the topographical shape of the limbus might not correspond to the color transition from cornea to sclera observed in en-face imaging techniques.

This study has some limitations. To accurately demarcate the limbus and estimate its shape, the coverage of the corneo-scleral area plays a key role in the measurement process. Some subjects, above all elderly ones, experienced difficulties to widely open their eyes, so they could not be included in our analysis. To overcome this limitation in clinical practice it is possible to use limbus sectors while analyzing data.[23] However if this is the case, we will lose the corneo-scleral transition points in superior and/or inferior sectors, leading to a less accurate limbus shape demarcation.

Having the full knowledge of the exact position and shape of the topographical limbus might be helpful for surgeons, in surgical procedures such as in trabeculectomy and non-penetrating deep sclerectomy; also in cataract and refractive surgery to assess the size of some types of intraocular lens3 or during limbal relaxing incisions procedure for the treatment of corneal astigmatism,[30] or even in the surgical technique of limbal transplantation. Also, in femtosecond laser–assisted cataract surgery the limbus, the pupil and the center of vision are used as anchor points in order to facilitate the surgery. Hence, the information about the real location of limbus, which is usually not concentric with pupil, could result in better centration of the IOL. Beyond the operating room, having an accurate tool to measure the transition points between the cornea and the sclera might facilitate the diagnosis of certain ocular disorders2 and may also assist ophthalmic practitioners.

***What was known***

WTW corneal diameter is a parameter amply used in clinical practice for vision care professionals, however current definitions of WTW corneal diameter are inconsistent.

***What this paper adds***

- Introduces the concept of the topographical limbus, which is a well-defined feature of the anterior eye surface.

- WTW corneal diameter and topographical limbus position are not equivalent. Normal topographical limbus is not symmetrical. In addition, on average the horizontal diameter is larger than the vertical one.

- Limbus shape is subject dependent but there are no age-related differences or differences between refractive groups.